# Mechanical Deformation of Nanoscale Metal Rods: When size and shape matters

M.J. Lagos[1,2], F. Sato[3], D.S. Galvão[2], D. Ugarte [1,2]*

[1]Laboratório Nacional de Luz Síncrotron-LNLS, C.P. 6192,

13083-970 Campinas SP, Brazil.

[2]Instituto de Física "Gleb Wataghin", Universidade Estadual de Campinas- UNICAMP,

13083-970 Campinas SP, Brazil.

[3]Instituto de Ciências Exactas, Universidade Federal de Juiz de Fora- UFJF,

13083-970 Campinas SP, Brazil.

*To whom correspondence should be addressed. E-mail: dmugarte@ifi.unicamp.br

Understanding nanomechanical response of materials represents a scientific challenge. Here, we have used in-situ electron microscopy to reveal drastic for the first time changes of structural behavior during deformation of 1-nm-wide metal rods as a function of temperature. At 300 K, stretched nanowires stay defect-free, while at 150 K, elongation is associated with planar defects. As size is reduced, energy barriers become so small that ambient thermal energy is sufficient to overcome them. Nanorods display an elastic regime until a mechanism with high enough blocking barrier can be nucleated. *Ab-initio* calculations revealed that contribution from surface steps overrule stacking fault energetics in nanorods, in such a way that system size and shape determines preferred fault gliding directions. This induces anisotropic behavior and, even large differences in elastic or plastic response for elongation or compression. These results provide a new framework to improve theoretical models and atomic potentials to describe the mechanical properties at nanoscale.

The mechanical properties of a strained nanoscale volume of matter represent a fundamental issue for understanding phenomena such as friction, fracture, adhesion, etc. Miniaturization levels are raising the need of accurately characterizing nanodevices and nanomaterials, to develop models and predictions of their mechanical performance and reliability[1]. But, we must also consider that the deformation of macroscopic matter continues to be a very interesting and dynamic research field displaying open questions and polemical issues.

The mechanical behavior of face centered cubic (FCC) metals is one of the most deeply studied cases, where plastic deformation can be understood on the basis of partial dislocation (PD) nucleation and propagation[2]. Following the Rice approach [3,4], the energetic related to PD formation can be described by the Generalized Stacking Fault surface (GSF) that indicates the interplanar energy originated by rigidly displacing one half of a fcc crystal along a [1 1 -2] direction on a (1 1 1) plane. A typical GSF curve has two bumps; to nucleate a PD the system must overcome an energy barrier, usually named unstable stacking fault energy ($\gamma_{usf}$). Once nucleated, the PD will generate a stable stacking fault (SF) of surface energy $\gamma_{ssf}$ (valley of GSF). The barrier to generate the trailing partial is associated with overcoming the second "bump" (barrier $\gamma_{usf} - \gamma_{ssf}$), when a total dislocation (TD) is produced. This model can not be extended to nanostructures, because size effects and surface energy may play a significant role.

The present computational power has allowed the realization of complex simulations in order to gather insight into nanoscale deformation mechanisms[1,5,6]. Nonetheless, time scales are very different from experiments[7], rendering unfeasible the analysis of temperature dependent effects and rate limited processes[6,8]. In consequence, experimental validation of models is seriously hindered and, it has become imperative to get experimental data analyzing the tensile deformation in individual nanosystems by means of time-resolved and atomic resolution imaging[1]. Here, we present a *in-situ* high resolution transmission electron microscopy (HRTEM) study of temperature effects on the structural evolution of stretched nm-wide metal rods. The nanorods are generated following the procedure introduced by Kondo and Takayanagi[9] (see a detailed description in Supporting Information). Firstly, holes are opened in a self-supported gold film by focusing the electron beam and, nanometric bridges are formed between holes. These constrictions spontaneously evolve, elongate and break; the dynamical process is video recorded (TV rate).

Fig. 1a shows some images of a 0.8 nm wide rod-like Au nanowire (NW) being elongated at room temperature (~300 K) along [1 1 0] direction (hereafter noted [1 1 0] wire). For an FCC metal stretched along the [1 1 0] direction, SFs should be generated by gliding {1 1 1} and {1 1 -1} planes[2] (see Fig. 2a). However, the NW stretched at 300 K seems to stay defect free and, in addition, the pillar-like wire breaks abruptly below a certain size[10] (see Supporting Information, Video 01). At low temperatures (~150 K), the NW structural evolution is significantly modified with formation of SFs and twins (TWs) (Figs. 1b,c; Supporting Information, Video02 and Video03). Also, the rod-like wires form bi-pyramidal junctions before rupture, leading to the formation of suspended atom chains[11,12]. In Figs. 1b,c, the [1 1 0] NWs are observed along a different direction and, the {1 1 1} slip planes appear horizontal in the images. Hence, defects associated with gliding these planes can be easily identified (see schematic draw in Fig. 2b). As expected, the SFs in Fig. 1b,c correspond to slip of the whole {111} planes[13] by a distance equivalent to a PD generation (Fig. 2b). When several SF faults are formed along the wire and, the minimal distance between them correspond to three hexagonal-compact-planes following FCC stacking (ABC, Fig. 1c).

The procedure used here to generate nanowires does not allow either to control the deformation of the nanowires or to get information of applied forces. However, experiments at different temperatures can be indeed realized and, more significantly, with a reasonable high efficiency (numerous events measured by high quality images of atomistic details). This allows a reliable statistical analysis of nanowire elongation and rupture processes. For example, numerous experiments were considered in the reported research, several hundred observations were considered for wires elongated at room temperature and more than one hundred for similar experiments realized at low temperature (for example, Fig. 1d shows several examples of wires, ~1 nm wide, containing planar defects when elongated at 150 K). In this way, the ensemble of analyzed experiments includes slightly different elongation speeds, different shear component contributions, apexes atomic structure, etc. Without going into further detail, a clear fact is revealed for the first time when comparing room and low temperature experiments: a defect-free elongation of 1-nm-wide gold wires at room temperature and, the occurrence of planar defects at low temperatures for wires of identical size at similar elongation speed.

We must emphasize that our results have clearly distinguished two distinct structural behaviors of 1–nm-wide gold wires. These two differentiated behaviors have been revealed by time resolved atomic resolution TEM experiments realized at different temperatures. Modern electron microscopy sample holders have attained a high degree of complexity. At present, they allow the controlled manipulation of metal nanowires (elongation, compression, shear, deformation rate) and, the measurement of applied forces, as well as electrical properties[11,14-18]. Unfortunately, it is only possible to make this kind of studies at room temperature. However, this kind of sample holder represents a very powerful tool. They may be used to analyze the size effect reported here by means of an extensive study at the single temperature (300 K) and, in addition the role of elongation/compression/shear deformation and its speed can also be discriminated and studied in detail.

Using the crystallographic Wulff method[19,20], it is possible to interpret the HRTEM images in detail (see Fig. 2) and, to demonstrate that the wires observed in Figs. 1a and 1b have exactly the same size and morphology (see Supporting Information). This allows a direct comparison between experiments realized at different temperatures. At ~300 K and within our time resolution (33 ms), the NW seems to easily annihilate defects during deformation; in contrast, defects are observable at ~150 K. This suggests that for Au rod of NM size thermal energy plays an essential role for quick defect recombination. Applying the Arhemius method, we can estimate the energy barrier ($\Delta E$) associated with SF recombination. Unfortunately, due to experimental constraints, only two different temperatures are available as input data (see discussion in Supporting Information); analyzing the experimentally observed SF lifetimes, a barrier $\Delta E \sim 40$ meV is roughly estimated (see Supplementary Information and Fig. S1). This accounts for experimental results where the nanorod SFs remain blocked a reasonable time at 150 K (3x thermal energy $\sim 39$ meV $\sim \Delta E$), and show a quick recombination at 300 K.

To further test this interpretation, we can perform experiments using another FCC metal with different SF barrier height. A point in case is Pt, because its macroscopic GSF barrier is about twice the gold value[21,22]. Then, we must also expect to observe SF defects during the elongation of Pt wires even if temperature is doubled (150 K → 300 K). This is in fact confirmed by room temperature deformation experiment of a [1 1 0] Pt NW of similar size (Fig. 3a). Extending even more the present concept, we should observe other defects with lower barriers at 150 K. Fig. 3b shows a [110] Pt nanorod containing a grain boundary moving along the wire (this Pt rod structure is equivalent to the Au rods in Fig. 1a,b; see Fig. 2b). A detailed analysis reveals the astonishing fact that a boundary dislocation can exist at the center of the pillar-like wire that is less that one nm in diameter (marked with an arrow in the figure).

In order to get further insights into the energetics of these processes, we have used *ab initio* calculations[23] to derive the total energy changes associated with the SF generation in the NWs shown in Fig. 2. This is equivalent to calculate a GSF surface for a macroscopic system[3,4,21,22], but we are intrinsically incorporating the surface contribution. For the metal rods studied in this work (Fig. 2), the {1 1 1} gliding planes display an elongated hexagonal shape (see Fig. 4a). We have considered the slip corresponding to a total dislocation movement generated by two successive PD shifts[3]. Six possible TD final positions are possible; considering the nanorod morphology, they can be grouped into three dissimilar structural configurations (indicated 1,2,3 in Fig. 4a,b; pathways 2 and 3 share the first PD shift). Fig. 4c display the total energy results for these TD paths. For macroscopic materials, the GSF surfaces come back to the baseline level after the TD generation (stacking sequence is recovered). In contrast, nanorod curves are superposed on high background attributed to the generation of surface steps. It can easily be realized that the energy variation related to SFs (bumps height) seem to be significantly less important than the surface contribution. Also, the surface term is always increasing; then, it represents a driving force trying to spontaneously annihilate the SFs.

A quick analysis of total energy curves reveal that the paths show quite different energy dependence. From a geometrical point of view, path 1 should be the preferred deformation mechanism induced a tensile force applied on the wire (the shift follows the projected shear force component parallel to the gliding plane). However, this path exposes large surface steps (schematically presented in Fig. 4b) that represent quite a high energy cost; for example the surface contribution may attain ~8 eV for the final TD stage. Path 2 requires an additional shear force (~ex. in a perpendicular direction), but the energy cost to form the $1^{st}$ PD is much lower because the induced surface step is much smaller (see Fig. 4b,c). This situation changes significantly for the $2^{nd}$ PD generation; the generated step is much larger, generating a significant energy increase (~7 eV). Concerning path 3, in spite of its final lower energy cost (and smaller surface step), it can not be considered as a stretching mechanism because it induces a shift opposite to the tensile effort. We must emphasize that these total energy results indicate that the $1^{st}$ PD of path 2 and 3 should be the preferred SF defect for the nanorods (step # 10 in Fig. 4c). This conclusion agrees with the experiments where observed SFs correspond to the formation of this particular configuration (schema in Fig 2b and experimental observation in Fig. 1b 30.4s). Finally, path 3 should be the preferred deformation mechanism form the energetic point of view, but it can only be active during compression. Then, the comparison of curves 1 and 3 reveals how surface contribution induces a clear anisotropy and, a quite different response is predicted for the mechanical properties during extension and compression. This phenomenon points out the essential role of size and shape of a nanosystem for the mechanical properties at nanoscale.

Our results provide the first direct quantitative experimental information and energetic understanding for defects generation in nanoscale metal rods. The relevant parameters to analyze

Lagos *et al* 7nanoscale deformation mechanisms are surface energy, morphology (determines how a defect will influence surface energy), defect blocking energy barrier (depends on material and systems size) and, finally the available thermal energy. For each kind of defect, material and temperature there is a threshold size for a particular defect to be considered an active nanoscale deformation mechanism. A nanosystem will behave as elastic until a defect with high enough blocking barrier can be nucleated. This study provides fundamental quantitative data to improve models and atomic potentials used in future temperature dependent simulations of the mechanical properties of nanostructures.

**References**

1. Alloca, C.; Smith, D. *Instrumentation and Metrology for Nanotechnology, Report of the National Nanotechnology Initiative*, Ch. 3 (available from www.nano.gov, 2005).

2. Nabarro, F. R. N. *Theory of Crystal Dislocations*, Dover, New York, 1987.

3. Rice, J. R. *J. Mech. Phys. Solids* **1992**, *40*, 239.

4. Tadmor, E. B.; Hai, S. *J. Mech. Phys. Solids* **2003**, *51*, 765.

5. Buehler, J. M. *Atomistic Modeling of Materials Failure*, Springer, New York, 2008.

6. MRS Bulletin, March 2009, *Atomistic Simulations of Mechanics of Nanostructures* Guest Editors: Hanchen Huang and Helena Van Swygenhoven (March 2009 Issue).

7. Agraït, N.; Yeyati, A. L.; van Ruitenbeek, J. M. *Phys. Rep.* **2003**, *377*, 81.

8. Warner, D. H.; Curtin, W. A.; Qu, S. *Nature Materials* **2007**, *6*, 876.

9. Kondo, Y.; Takayanagi, K. *Phys. Rev. Lett.* **1997**, *79*, 3455.

10. Rodrigues, V.; Fuhrer, T.; Ugarte, D. *Phys. Rev. Lett.* **2000**, *85*, 4124.

11. Onishi, H.; Kondo, Y.; Takayanagi, K. *Nature* **1998**, *395*, 780.

12. Yanson, A. I.; Bollinger, G. R.; van den Brom, H. E.; Agrait, N.; van Ruitenbeek, J. M. *Nature* **1998**, *395*, 783.

13. Landman U.; Luedtke, W. D.; Burnham, N. A.; Colton, R.J. *Science* **1990**, *248*, 454.

14. Kizuka, T *Phys. Rev. Lett.* **1998**, *81*, 4448.

15. Erts, D.; Olin, H; Ryen, L.; Olsson, E.; Thölén, A. *Phys. Rev. B* **2000**, *61*, 12725.

16. Peng, B.; Locascio, M.; Zapol, P.; Li, S.; Mielke, S.; Schatz, G.; Espinosa, H. D. *Nat. Nano.*, **2008**, *3*, 626.

17. Kizuka, T. *Phys. Rev. B* **2008**, *77*, 1155401.

18. Kurui, Y.; Oshima, Y.; Okamoto, M.; Takayanagi, K *Phys. Rev. B* **2009**, *79*, 165414.

19. Marks, L. D. *Rep. Prog. Phys*. 1**994**, *57*, 603.

20. Rodrigues, V.; Ugarte, D. *Nanowires and Nanobelts Vol. 1* Ch. 6, ed. Wang, Z.L., Kluwer Acad. Pub., Boston, 2003.

21. Mehl, M.J.; Papaconstantopoulos, D. A.; Kioussis, N.; Herbranson, M. *Phys. Rev. B* **2000**, *61,* 4894.

22. Van Swygenhoven, H.; Derlet, P. M.; Froseth, A. G. *Nature Materials* **2004**, *3*, 399.

23. Sanchez-Portal, D.; Ordejón, P.; Artacho, E.; Soler, J. M. *Int. J. Quantum Chem.* **1997**, *65*, 453.

**Acknowledgements**

P.C. Silva and J. Bettini are acknowledged for assistance during HRTEM work. We thank V. Rodrigues for comments. Supported by LNLS, FAPESP and CNPq.

Correspondence and requests for materials should be addressed to D.U.

**Supporting Information vailable:** Methods. Nanowire Generation. Interpretation of atomic resolution HRTEM images. DFT Calculations. Estimation of stacking fault energy barrier from HRTEM videos. Time evolution of atomistic processes during the elongation and rupture of a one-nm-wide Au rods at room and low temperature (150 K).

**Figure Captions.**

Figure 1. Representative electron microscopy snapshots showing the atomic details associated with the elongation of nanometric rod-like Au junctions along the [110] direction. a) stretching and rupture of a NW at ~300 K; note that the structure seems to stay straight and defect-free. b,c) tensile deformation process of nanometer wide junctions occurring at low temperature (~150 K); in this case, the formation of planar structural defects can be easily identified (SFs and TWs, indicated by arrows). d) Selected snapshots of gold nanowires stretched at 150 K and displaying the formation of stacking faults (marked with arrows). Atomic positions appear dark. The elongation rate of the wires were ~0.02, ~0.03 and ~0.02 nm/s for 1a, 1b and 1c, respectively.

Figure 2. a) Left side: schematic representation of the atomic arrangement of the [110] rod-like NW formed by 5 {200} atomic planes shown in Fig. 1a (0s); right side: proposed cross-section as deduced from the Wulff construction. b) Top: image interpretation of a wire formed by 4 {-11-1} planes as displayed in Fig 1b (0s), the stacking of hexagonal-close-packed planes is indicated (ABC sequence for a defect-free FCC crystal). b) Bottom: pictorial examples of structural changes that may be generated by gliding one rod section over the $4^{th}$ {111} atomic plane. We show at left, SF generated by partial dislocation formation (PD); the central draw shows a total dislocation movement (TD) that generates a higher surface step but realigns the remaining {-11-1} planes). Finally at right, the generation of a twin defect at the $5^{th}$ layer is exemplified.

Figure 3. Atomic resolution images of Pt nanorods stretched along the [110] direction; atomic positions appear dark. a) in contrast to Au systems, at room temperature (300K) Pt rods of similar size (4 {111} atomic plane wide) display the formation of SFs (arrowed in image 34s). b) the rod displays a grain boundary (arrowed in the NW upper region, observation at temperature 150 K), which is moving down along the wire (a black bar serves as a position reference to visualize the boundary movement). A careful analysis of (b) reveals a boundary dislocation at the center of the pillar-like wire (atomic planes indicated by the arrow).

Figure 4. Theoretical *ab initio* calculations of the total energy changes associated with SF generation in a nanorod as shown in Fig. 2 and, containing 106 atoms. The {111} slip planes have a hexagonal shape and are composed of 14 atoms (a); the sequential stacking of these planes is represented by different colours A (black, continuous line), B (white, dotted line). The upper part of the figure (a) indicate total dislocation shift of a plane in site B (initial position is marked 0) to other B-site following three different pathways (final positions marked 1, 2, 3; for each of this positions there is an equivalent path but moving downwards). The TD movements are generated by two successive PD movements with an intermediate stable SF position (C site, in grey). (b) Schema representing the PD (upper part) and TD (lower part) structural configuration for the different paths; hexagons surrounding each atomic plane (A or B) allow a quick evaluation of the exposed areas generated by the surface steps. (c) Total energy changes associated with the three different possible pathways.

# Figure 1

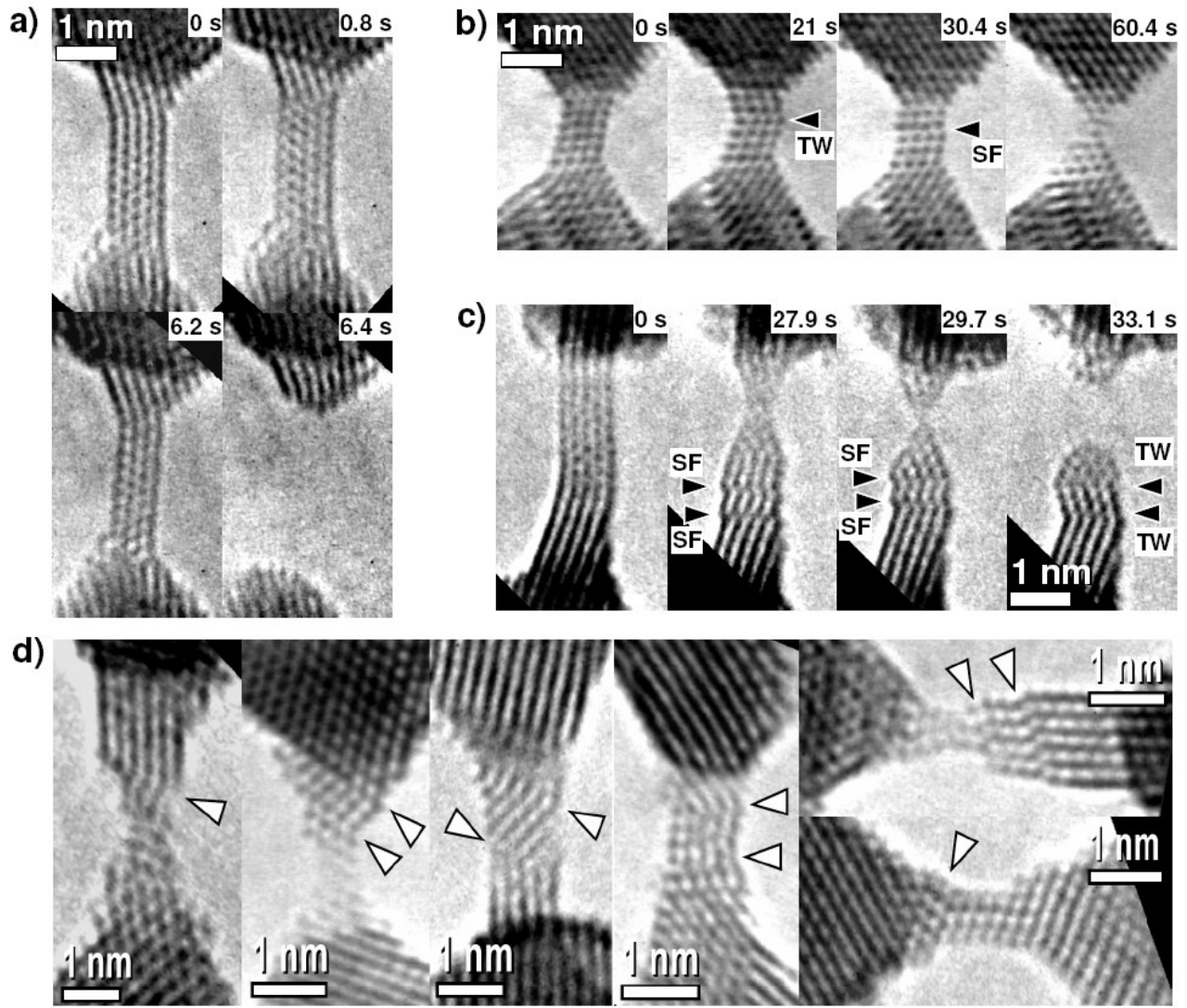

# Figure 2

# Figure3

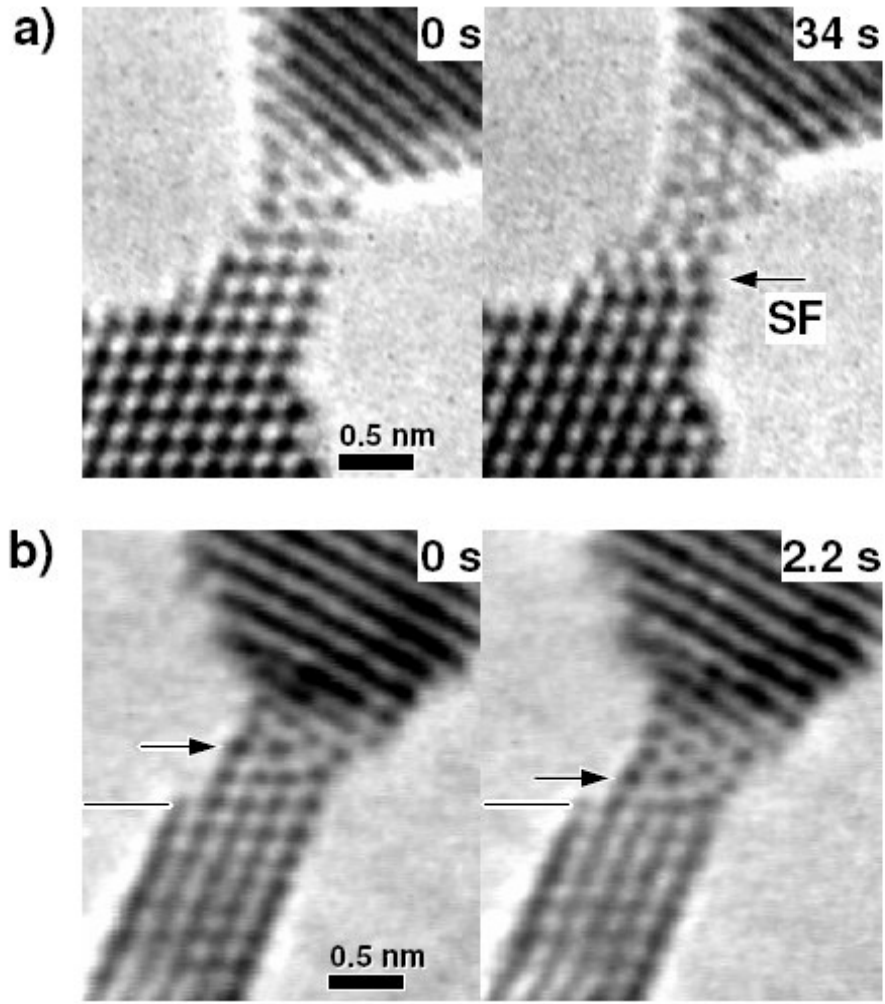

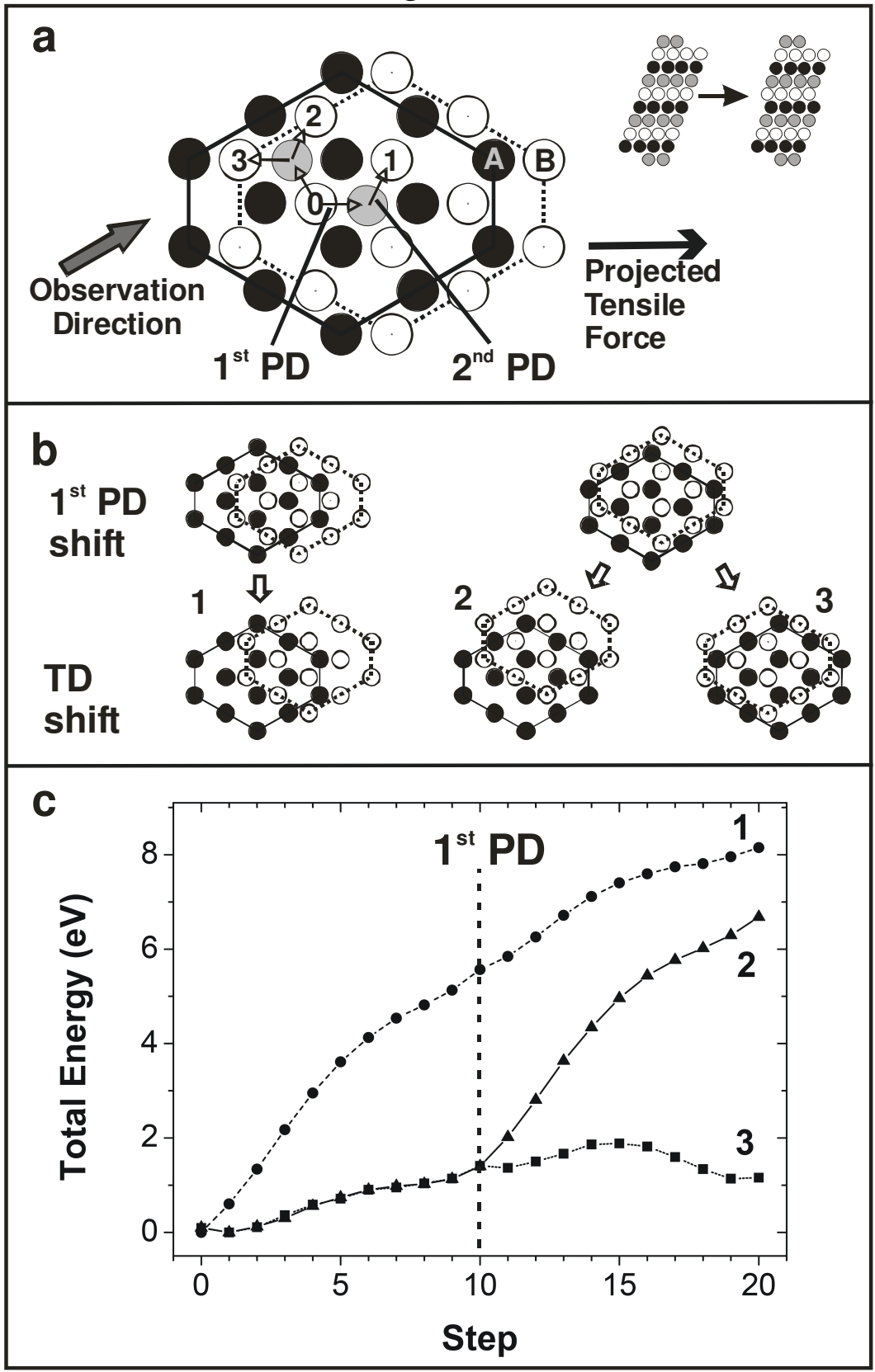

# Supporting Information

## Mechanical Deformation of Nanoscale Metal Rods:

## When size and shape matters

M.J. Lagos[1,2], F. Sato[3], D.S. Galvão[2], D. Ugarte [1,2]*

[1]Laboratório Nacional de Luz Síncrotron-LNLS, C.P. 6192,

13083-970 Campinas SP, Brazil.

[2]Instituto de Física "Gleb Wataghin", Universidade Estadual de Campinas- UNICAMP,

13083-970 Campinas SP, Brazil.

[3]Instituto de Ciências Exactas, Universidade Federal de Juiz de Fora- UFJF,

13083-970 Campinas SP, Brazil.

*To whom correspondence should be addressed. E-mail: <dmugarte@ifi.unicamp.br>

**This PDF File includes:**

Methods

    Nanowire Generation

    Interpretation of atomic resolution HRTEM images.

    Theoretical: DFT Calculations

    Estimation of stacking fault energy barrier in Au nanorods from HRTEM videos

Figure    S1

Video Captions:    Video01

                  Video02

                  Video03

**METHODS**

**Nanowire Generation**

Metal nanorods were produced inside the HRTEM (JEM-3010 URP 300kV, 0.17 nm point resolution). The dynamic HRTEM observations have been realized at room (~300 K) and low temperature (~150 K) using a liquid $N_2$ cooled Gatan 613-DH sample holder[2,3]. Videos were recorded with a high sensitive TV camera (Gatan 622SC, time resolution 33 ms). This method allows the time-resolved HRTEM imaging of NWs with a remarkable quality; however, we can neither control the elongation direction nor measure the applied force[2,4]. Typical elongation speed is about 0.1 nm/s[4]. In order to get the high quality atomic resolved images presented in this work, instrumental constrains lead to the fact that only two different temperatures are accessible ~300 and ~150 K. These temperatures correspond to samples holders that can get a reasonable mechanical stability by using a direct conductive cooling of the HRTEM sample by a stable thermal bath (ex. liquid nitrogen reservoir). Liquid He cooled holders maybe a third option, but unfortunately, they were not available and, usually vibrations hinder the generation of high quality atomic resolution images.

HRTEM images allow the direct measurement distances between atoms with sub angstrom precision; typical pixel size for time-resolved images is about 0.01 nm. To perform an elongation rate measurement, we chose a high contrast region that remains stable from each apex during the experiment. Using cross-correlation, we can measure movements of these regions with pixel accuracy in different frames. Cross-correlation methods are at present the standard technique for realigning sample position in modern analytical scanning electron microscopy softwares. By measuring the relative distance between these selected regions in each apex, we can measure modifications in the relative position between apexes. In this way the average elongation rate is measured from the experiments.

**Interpretation of atomic resolution images.**

In simple terms, HRTEM images yield just a bi-dimensional projection of the atomic structure. However, it is possible to deduce the three dimensional morphology of a nanoparticle or nanowire (NW) by means of the geometrical Wulff construction[4,5] that minimizes the surface energy by predicting the correct crystal faceting. For a rod-like Au wire elongated along the [110] direction that is formed by 5 {200} atomic-planes (0.8 nm) in width (see Fig. 1a, 0s) and observed along the [1-10] direction, the predicted cross-section has already been reported in Refs. 4 (see Figure 2a). Note that this rod cross-section can also be considered as formed by 4 stacked {-1 1 -1} planes parallel to the wire axes (see cross-section in Fig. 2a). The Au NWs shown in Figs. 1b,c show a different image contrast, because these wires are observed along a different crystallographic direction ([10-1]). In these conditions, the [110] wire axis is not parallel to the image plane and the {111} family of planes appears horizontal in the atomically resolved images. The rod-like wire in Fig. 1b appears formed by 4 {-1 1 -1} atomic planes, indication that this wire has exactly the same size and structure that the wire studied in Fig. 1a. The rod-like Pt wires displayed in Fig. 3a and in Fig. 3b (the upper part of the Pt wire) are also formed by the stacking of 4 {-1 1 -1} atomic plane in width. Then, considering that the surfaces energies of Pt and Au crystal surfaces are quite similar[6], we conclude that also their structure are identical. This allows the direct comparison of elongation experiments in similar NWs made of different materials (Au, Pt) and studied at different temperatures (300 K, 150 K).

**Theoretical: DFT Calculations**

We have carried out *ab initio* DFT calculations (SIESTA code[7] considering spin polarization, relativistic corrections, and Local Density Approximation (LDA)) to obtain the total energy profiles associated with the generation of a stacking fault in a NW. The geometric structures were obtained based on bulk atomic distances. For optimizing gold bulk parameters, the atoms positions and lattice parameters were set free to attain inter-atomic forces below 0.01 eV/Å. The displacement from a defect-free NW to a total dislocation position (Fig. 2b) was divided in 20 steps, considering the two PD shifts for each analyzed path. Each point used to build the nanorod glide-plane energy profile, was obtained by a single point energy calculation (one self-consistent field) of the electronic part and the nuclear positions were kept frozen.

**Estimation of stacking fault energy barrier in Au nanorods from HRTEM videos**

The estimation of the energy barrier ($\Delta E$) in a thermally activated process is obtained usually by means of the Arhemius plot where we relate the process occurrence rate ($R$) to the temperature. This plot assumes the rate function will be proportional to $\text{Exp}(-\Delta E/ k_BT)$ (where Exp is the exponential function, $k_B$ is Boltzmann constant, and $T$ the temperature). Plotting $ln(R)$ vs. $1/k_BT$, we should get a straight line whose slope directly yield $\Delta E$. The model above, can also be interpreted in terms of life-time ($\tau = 1/R$) of a state for each temperature. The time-resolved atomic resolution images of stretched Au nanowires allow the measurement of the time elapsed between SF (Stacking Fault) formation and annihilation. As the NW elongation speed is so low (~0.1 nm/s), we can in a first approximation think about a quasi-static study where we can neglect the role of the stress contribution to fault recombination. In these terms, we can assume that the SF recombination is purely due to thermal energy.

At ~150 K, SFs are observable during the video recording, then we have measured the time interval ($\Delta t$) between formation and annihilation of each SF observed in NWs of similar size (a total of 11 events). Continuing our simple thermodynamical modeling, for a fixed temperature ($T_1 = 150$ K) the probability of observing a certain time interval will be proportional to $\text{Exp}(-\Delta t/\tau_1)$ (where $\tau_1$ is the SF life-time at $T = T_1$). In consequence, a logarithmic plot of the number of occurrence events for each time interval, should be a straight line of slope ($-1/\tau_1$). Fig. S1 shows the logarithmic plot of experimental data that confirms the predicted linear behavior and, it allows a SF life-time estimation of $\tau \sim 1s$ at T~150 K. At ~300 K, no SF is observable in the video recording, then we may think that the SFs lifetime is of the order of our time resolution (0.033 s).

Experimental constrains limit the electron microscopy available data to merely two possible temperatures. Making the rough approach of using the only two available temperatures life-times as input data for the Arhemius plot, the derived barrier height estimation results in $\Delta E \sim 40$ meV for gold nanowires with the atomic structure displayed in Fig 2a. Although, the high difficulty of getting the time resolved HRTEM images for such small systems and the approximations used for the barrier derivation, we must emphasize that the $\Delta E$ value account for experimental observations. On the basis of bulk material knowledge, the barrier to be overcome should be $A.(\gamma_{usf}-\gamma_{ssf})$ (where $A$ is the stacking fault area)[8-11] ; this simple calculation yields a ~140 meV, much higher than derived from the HRTEM observations.

**References**

(1) Kondo, Y.; Takayanagi, K. *Phys. Rev. Lett.* **1997**, *79*, 3455.

(2) Oshima, Y.; Onga, A.; Takayanagi, K. *Phys. Rev. Lett.* **2003**, *91*, 205503.

(3) Lagos, M.; Rodrigues, V.; Ugarte, D. *J. Electron Spectrosc. Rel. Phenom.* **2007**, *156-158*, 20.

(4) Rodrigues, V.; Ugarte, D. *Nanowires and Nanobelts Vol. 1* Ch. 6, ed. Wang, Z.L., Kluwer Acad. Pub., Boston, 2003.

(5) Marks, L. D. *Rep. Prog. Phys*. 1**994**, *57*, 603.

(6) Rodrigues, V., Sato, F. , Galvão, D. S.; Ugarte, D. *Phys. Rev. Lett.* **2007**, *99*, 255501.

(7) Sanchez-Portal, D.; Ordejón, P.; Artacho, E.; Soler, J. M. *Int. J. Quantum Chem.* **1997**, *65*, 453.

(8) Rice, J. R. *J. Mech. Phys. Solids* **1992**, *40*, 239.

(9) Tadmor, E. B.; Hai, S. *J. Mech. Phys. Solids* **2003**, *51*, 765.

(10) Mehl, M.J.; Papaconstantopoulos, D. A.; Kioussis, N.; Herbranson, M. *Phys. Rev. B* **2000**, *61,* 4894.

(11) Van Swygenhoven, H.; Derlet, P. M.; Froseth, A. G. *Nature Materials* **2004**, *3*, 399.

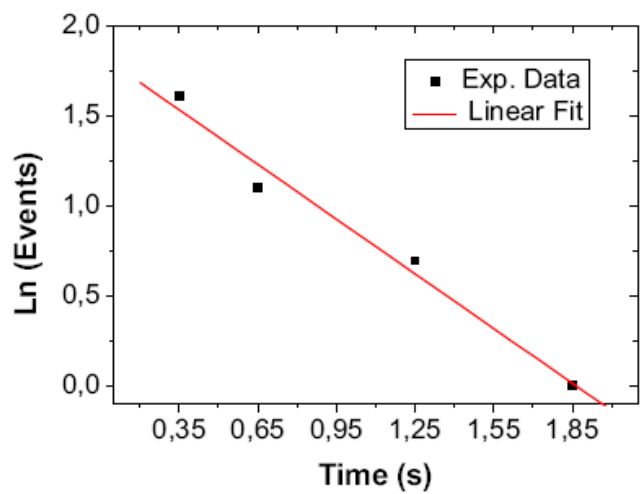

Figure S1. Experimental data of time intervals ($\Delta t$) between SF formation and annihilation in Au nanowires at 150 K. The logarithmic plot represents the number of events that a time interval is observed vs. time interval. A clear linear behavior is observed indicating a well-defined exponential dependence. A linear fit allows the derivation of the SF life-time at this temperature.

**Video Captions:**

Video 01:    Elongation and rupture of a rod-like Au NW under tensile stress along the [110] crystalline direction at room temperature (~300 K).  The observation direction is ~[1-10], the wire width is ~0.8 nm (or 5 {200} atomic planes, $d_{200}$ ~0.2 nm).

Video 02:    Elongation and rupture of a rod-like Au NW under tensile stress along the [110] crystalline direction at low temperature (~150 K).  From Wulff´s construction rules, the cross-section of this wire should be identical to the NW displayed in Video01. The observation direction is ~[10-1], the wire width is ~0.7 nm (or 4 {-11-1} atomic planes, $d_{111}$ ~0.23 nm).

Video 03:  Elongation and rupture of a slightly thicker (~0.92 nm, or 5 {-11-1} atomic planes) rod-like Au NW under tensile stress along the [110] crystalline direction at low temperature (~150 K).